\documentclass{article}
\usepackage{geometry} 
\usepackage{amsmath,amsfonts,amsthm,amssymb,amscd}
\usepackage{accents}
\usepackage{latexsym}
\usepackage{hyperref}
\usepackage{graphicx}
\usepackage{cite}
\usepackage{authblk}
\usepackage{bmpsize}
\usepackage{caption}
\usepackage[all]{xy}
\usepackage{url}

\hypersetup{
setpagesize=false,
 bookmarksnumbered=true,%
 bookmarksopen=true,%
 colorlinks=true,%
 linkcolor=blue,
 citecolor=red,
 urlcolor=blue
}

\makeatletter
 
  \@addtoreset{equation}{section}
\makeatother

\title{Gravitational waves in Modified Gauss-Bonnet Gravity }

\author[1,2,3]{Tomohiro Inagaki \thanks{inagaki@hiroshima-u.ac.jp}}
\author[4]{Masahiko Taniguchi \thanks{masa-taniguchi@hiroshima-u.ac.jp}}

\affil[1]{\it{Information Media Center, Hiroshima University, Higashi-Hiroshima, 739-8521,
Japan}}
\affil[2]{\it{Core of Research for the Energetic Universe, Hiroshima University,
Higashi-Hiroshima, 739-8526, Japan}}
\affil[3]{\it{Lab. Theor. Cosmology, Tomsk State University of Control Systems and
Radioelectronics (TUSUR), 634050 Tomsk, Russia}}
\affil[4]{\it{Department of Physics, Hiroshima University, Higashi-Hiroshima, 739-8526, Japan}}

\date{}

\begin{document}
\maketitle

\begin{abstract}
We study the gravitational waves in modified Gauss-Bonnet gravity.
Applying the metric perturbation around a cosmological background, we obtain explicit expressions for the wave equations.
It is shown that the speed of the traceless mode is equal to the speed of light.
An additional massive scalar mode appears in the propagation of the gravitational waves.
To find phenomena beyond the general relativity the scalar mode mass is calculated as a function of the background curvature in some typical models.
\end{abstract}

\section{Introduction}
In 2015, LIGO first detected gravitational waves (GWs) from the merger of a binary black holes of around 36 and 29 solar masses, and its result is consistent with the prediction of General Relativity (GR)~\cite{Abbott:2016blz}. 
It indicates that GR is still correct under strong gravity. However, astrophysical observations provide phenomena which can not be explained in GR, like the accelerating expansion. 
It is considered that the cosmic expansion has two phases. 
The one is Inflation in the very early universe which is proposed to solve the horizon and flatness problems~\cite{Sato:1980yn,Guth:1980zm}. 
The other is the current expansion of the universe from the observational consequence of type Ia supernovae~\cite{Riess:1998cb}, cosmic microwave background~\cite{Samtleben:2007zz}, baryon acoustic oscillations~\cite{Bassett:2009mm} and so on. 

Modified gravity is one of candidates to induce the cosmic expansion at very early universe and the current period~\cite{Nojiri:2017ncd}. 
It is shown that the existence of the modified gravity which pass the inspections from the observations. The most popular model of the modified gravity is $F(R)$ gravity which replace Ricci scalar, $R$, in the Einstein-Hilbert action by an arbitrary function of Ricci scalar, $F(R)$. 
Since the gravitational degree of freedom in $F(R)$ gravity is three~\cite{Myung:2016zdl,Gong:2017bru,Moretti:2019yhs}, GWs in $F(R)$ gravity have, in addition to the tensor modes propagation, a scalar mode propagation. 
The wave equations for $F(R)$ gravity in a cosmological background is given by~\cite{Katsuragawa:2019uto}
\begin{equation}\label{eq:mass of F(R)}
[\square-m_{F(R)}^2]\delta \Phi=0, \quad m_{F(R)}^2= \frac{1}{3}\left(\frac{F'(\tilde{R})}{F''(\tilde{R})}-\tilde{R}\right),
\end{equation}
where $\delta \Phi$ indicates the fluctuation of the scalar mode. The mass, $m_{F(R)}$ depends on the background curvature. The non-vanishing mass means that the speed of the scalar mode propagation is less than the light speed and it constrains the $F(R)$ gravity models~\cite{Katsuragawa:2019uto,Yang:2011cp,Vainio:2016qas,Gogoi:2019zaz,Sharif:2017ahw}. 
The tensor modes in $F(R)$ gravity are massless and propagate with the light speed~\cite{Gong:2017bru,Yang:2011cp}.
It has been pointed out that the modified gravity changes the graviton amplitude and the propagation phase~\cite{Capozziello:2017vdi,Nojiri:2017hai}.

$F(\mathcal{G})$ gravity is an alternative model of the modified gravity which is proposed in Ref.~\cite{Nojiri:2005jg}.
The topological Gauss-Bonnet (GB) invariant, $\mathcal{G}$, which contains the contractions of Ricci and Riemann tensors is derived from string theory at high energy as a low energy effective action~\cite{Boulware:1985wk}. A scalar field coupled with the GB invariant is introduced in the Einstein-GB gravity.
In the Einstein-GB gravity the gravitational wave speed constrains the scalar coupling to the GB invariant~\cite{Odintsov:2019clh}.
In $F(\mathcal{G})$ gravity an arbitrary function of the GB invariant is included in the Einstein-Hilbert action and applied some cosmological problems~\cite{Li:2007jm,Nojiri:2010wj}. 
From the topological property of GB invariant, the equation of motion in $F(\mathcal{G})$ gravity only has 2nd derivative terms. It is shown the degrees of freedom in $F(\mathcal{G})$ gravity is also three~\cite{Astashenok:2015haa}, and the tensor modes are massless. 
The model can be generalized to $F(R,\mathcal{G})$ gravity in which the Lagrangian contains an arbitrary function with respect to $R$ and $\mathcal{G}$~\cite{Nojiri:2006ri,DeLaurentis:2015fea}. $F(R)$ and $F(\mathcal{G})$ gravity can be described as special cases of $F(R,\mathcal{G})$ gravity. The degrees of freedom in $F(R,\mathcal{G})$ is four. Remarkably, the degrees of freedom reduce to three in Friedmann-Lemaître-Robertson-Walker(FLRW) background~\cite{DeFelice:2010hg}. 

In this paper the scalar mode in $F(\mathcal{G})$ and $F(R,\mathcal{G})$ gravity are investigated in De Sitter background. The experiments to measure the polarizations of GWs have already been started in Advanced LIGO and Virgo detectors~\cite{Abbott:2018utx}. At the moment, it is hard to discern the polarizations of GWs but the additional detectors will work in the near future, KAGRA~\cite{Aso:2013eba} and LIGO-India~\cite{Unnikrishnan:2013qwa}, LISA~\cite{Danzmann:2003tv}, DECIGO~\cite{Kawamura:2011zz}. It is expected to test the models of $F(\mathcal{G})$ gravity and $F(R,\mathcal{G})$ in the precise polarization measurements.

This paper is organized as follow. In Sec.~\ref{GWs in F(G)}, we briefly review the equation of motion in $F(\mathcal{G})$ gravity. Using the perturbations around cosmological background, we obtain the wave equations of the trace modes and scalar mode. In Sec.~\ref{GWs in F(R,G)}, we calculate the wave equation of the scalar mode in $F(R,\mathcal{G})$ gravity. 
In Sec.~\ref{Calculation}, the scalar mode mass is calculated in some typical models of $F(R)$ and $F(\mathcal{G})$ gravity. In Sec.~\ref{observations}, we discuss the possibility to detect the scalar mode propagation in the future GWs observations. Some concluding remarks are given in Sec.~\ref{conclusion}.

\section{GWs in $F(\mathcal{G})$ gravity\label{GWs in F(G)}}
\subsection{ $F(\mathcal{G})$ gravity}
$F(\mathcal{G})$ gravity is defined by the action~\cite{Nojiri:2005jg}
\begin{equation}\label{eq:action F(G)}
S_{G B}=\int d^{4} x \sqrt{-g}\left(\frac{M_{\rm pl}^2}{2}R+F(\mathcal{G})+\mathcal{L}_{\text {matter}}\right),
\end{equation}
where $M_{\rm pl}$ denotes Planck mass and $\mathcal{G}$ is the GB invariant, $\mathcal{G} = R^2-4R^{\mu\nu}R_{\mu\nu}+R^{\lambda\mu\rho\nu}R_{\lambda\mu\rho\nu}$. 
A key difference between $F(\mathcal{G})$ and $F(R)$ gravity comes from contractions of Ricci and Riemann tensors.
The GB invariant has topological property known as Chern-Gauss-Bonnet theorem~\cite{Chern:1944},
\begin{equation}\nonumber
\int d^4x \sqrt{-g}\mathcal{G} =8\pi^2 \chi (M), 
\end{equation}
where $ \chi(M)$ is the Euler characteristic. 
The variation of the action (\ref{eq:action F(G)}) with respect to the metric tensor yields the equation of motion (EoM).
In the vacuum the contribution from the matter Lagrangian, $\mathcal{L}_{\text {matter}}$, is dropped and EoM is found to be
\begin{align}\label{eq:EoM in F(G)}\nonumber
&\frac{M_{\rm pl}^2}{2}\left(-R_{\mu\nu}+\frac{1}{2}g_{\mu\nu}R\right)
+\frac{1}{2}g_{\mu\nu}F(\mathcal{G})-\frac{1}{2}g_{\mu\nu}\mathcal{G}F'(\mathcal{G})-2g_{\mu\nu}R\square F'(\mathcal{G})
\\ \nonumber
&+2R\nabla_\mu\nabla_\nu F'(\mathcal{G})+4R_{\mu\nu}\square F'(\mathcal{G})+4g_{\mu\nu}R^{\rho\sigma}\nabla_\rho\nabla_\sigma F'(\mathcal{G})
\\ 
&-4{R^\rho}_{\nu}\nabla_\rho\nabla_\mu F'(\mathcal{G})-4{R^\rho}_\mu\nabla_\rho\nabla_\nu F'(\mathcal{G})-4{{{R_\mu}^\rho}_\nu}^\sigma\nabla_\rho\nabla_\sigma F'(\mathcal{G})=0.
\end{align}
Eq.~(\ref{eq:EoM in F(G)}) does not contain the first derivative of the curvature, $R$, and the Gauss-Bonnet invariant, $\mathcal{G}$.
As is known, the Einstein equations of GR are obtained for Gauss-Bonnet gravity, $F(\mathcal{G})=\mathcal{G}$.

\subsection{Wave equations of $F(\mathcal{G})$ gravity}

To find the wave equation of $F(\mathcal{G})$ gravity in De Sitter background
we employ the metric perturbations around the background,
\begin{equation}\label{eq:perturbations for metrics}
g_{\mu\nu}=\tilde{g}_{\mu\nu}+h_{\mu\nu},
\end{equation}
where $\tilde{g}_{\mu\nu}$ shows the background metric.
The Gauss-Bonnet invariant is perturbatively expanded as 
\begin{align}\label{eq:perturbation of GB invariant}
\mathcal{G}\simeq \tilde{\mathcal{G}}+\delta\mathcal{G}+\mathcal{O}(h^2).
\end{align}
Under the De Sitter background the Riemann tensor and Ricci tensor can be written 
by the background metric, $\tilde{g}_{\mu\nu}$, and the background scalar curvature, $\tilde{R}$,
\begin{align} 
\tilde{R}_{\mu\rho\nu\sigma}&=\frac{\tilde{g}_{\mu\nu}\tilde{g}_{\rho\sigma}-\tilde{g}_{\mu\sigma}\tilde{g}_{\nu\rho}}{12}
\tilde{R}, \label{eq:Riemann scalar for de Sitter} \\
\tilde{R}_{\mu\nu}&=\frac{\tilde{g}_{\mu\nu}}{4}\tilde{R}.  \label{eq:Ricci tensor for de Sitter}
\end{align}
From Eqs.~(\ref{eq:Riemann scalar for de Sitter}) and (\ref{eq:Ricci tensor for de Sitter}), the background Gauss-Bonnet invariant is given by
\begin{align}\label{eqs:backgroud of GB}
\tilde{\mathcal{G}}=\frac{\tilde{R}^2}{6}.
\end{align}
For the background metric Eq.~(\ref{eq:EoM in F(G)}) is simplifies to
\begin{align*}
\frac{M_{\rm pl}^2}{2}\tilde{R}+2F(\tilde{\mathcal{G}})-2\tilde{\mathcal{G}}F'(\tilde{\mathcal{G}})=0.
\end{align*}

Next we calculate the equation of motion up to the first order of the metric perturbations.
The Riemann, Ricci tensors and Ricci scalar are expanded as
\begin{align}
R_{\mu\rho\nu\sigma}&\simeq \tilde{R}_{\mu\rho\nu\sigma}+\delta R_{\mu\rho\nu\sigma}+\mathcal{O}(h^2), \label{eq:perturb Riemann tensor}
\\
R_{\mu\nu}&\simeq\tilde{R}_{\mu\nu}+\delta R_{\mu\nu}+\mathcal{O}(h^2), \label{eq:perturb Ricci tensor} 
\\
R&\simeq \tilde{R}+\delta R +\mathcal{O}(h^2). \label{eq:perturb Ricci scalar}
\end{align}
The perturbative terms can be expressed in terms of $h_{\mu\nu}$~\cite{Weinberg:2008zzc},
\begin{align}
\delta R_{\mu\nu}&=-\frac{1}{2}(\square h_{\mu\nu}+\nabla_\mu\nabla_\nu h -\nabla_\mu\nabla^\lambda h_{\lambda\nu}-\nabla_\nu\nabla^{\lambda}h_{\lambda\mu}-\frac{2}{3}\tilde{R}h_{\mu\nu}+\frac{\tilde{R}}{6}\tilde{g}_{\mu\nu}h),
\\
\delta R&=-\square h +\nabla^\mu\nabla^\nu h_{\mu\nu}-\frac{\tilde{R}}{4}h,
\end{align}
where $\nabla_\mu$ represents the covariant derivative in De Sitter metrics and  $\square$ is D'Alembert operator, $\square \equiv \tilde{g}^{\mu\nu}\nabla_\mu\nabla_\nu$.
The perturbations of Gauss-Bonnet invariant is found to be
\begin{align}\label{eq:perturbation of gauss-bonnet}
\delta \mathcal{G}= \frac{\tilde{R}}{3}\left(-2\tilde{R}^{\mu\nu}h_{\mu\nu}+\tilde{g}^{\mu\nu}\tilde{g}^{\alpha\beta}\delta R_{\alpha\mu\beta\nu} \right)
=\frac{\tilde{R}}{3}\delta R,
\end{align}
where we use the following expression of the perturbations of Ricci scalar,
\begin{align}\label{eq:Curvature with respect to Ricci scalar}
\delta R=-\tilde{R}^{\mu\nu}h_{\mu\nu}+\tilde{g}^{\mu\nu}\delta R_{\mu\nu}
=-2\tilde{R}^{\mu\nu}h_{\mu\nu}+\tilde{g}^{\mu\nu}\tilde{g}^{\alpha\beta}\delta R_{\alpha\mu\beta\nu}.
\end{align}
$F(\mathcal{G})$ and $F'(\mathcal{G})$ are expanded as
\begin{align}
&F(\mathcal{G})\simeq F(\tilde{\mathcal{G}})+F'(\tilde{\mathcal{G}})\delta \mathcal{G},
\\
&F'(\mathcal{G})\simeq F'(\tilde{\mathcal{G}})+F''(\tilde{\mathcal{G}})\delta \mathcal{G}. 
\end{align} 
Therefore the equation of motion (\ref{eq:EoM in F(G)}) reduces to
 \begin{equation}\label{eq:perturbed EoM in F(G)}
-\frac{M_{\rm pl}^2}{2}\left[\delta R_{\mu\nu}-\frac{1}{2}\tilde{g}_{\mu\nu}\delta R-\frac{1}{4}\tilde{R}h_{\mu\nu} \right]+\frac{1}{3}\left[\tilde{g}_{\mu\nu}\square - \nabla_\mu\nabla_\nu +\tilde{g}_{\mu\nu}\frac{\tilde{R}}{4} \right]F''(\tilde{\mathcal{G}})\delta\mathcal{G} =0.
\end{equation}

To find a wave equation for the physical degrees of freedom we take the following gauge conditions,
\begin{equation}
\nabla^\mu \bar{h}_{\mu\nu}=0,\qquad \bar{h}_{0i}=0 \quad(i=1,2,3),
\end{equation}
where $\bar{h}_{\mu\nu}$ is defined as 
\begin{equation}
\bar{h}_{\mu \nu}=h_{\mu \nu}-\frac{1}{2} \tilde{g}_{\mu \nu} h.
\end{equation}
It should be noted that the traceless of the metric perturbation is also imposed as the gauge condition in GR.
A discrepancy between GR and $F(\mathcal{G})$ gravity is found by dividing the metric perturbation $h_{\mu\nu}$ into the 
traceless and scalar parts,
\begin{equation}\label{eq:div:traceless}
h_{\mu\nu}=h^T_{\mu\nu}+\frac{h}{4}\tilde{g}_{\mu\nu},
\end{equation}
where the traceless part, $h_{\mu\nu}^T$, holds
\begin{equation}
\tilde{g^{\mu\nu}}h_{\mu\nu}^T=0, \qquad h^T_{0i}=0\quad(i=1,2,3).
\end{equation}
Substituting Eq.~(\ref{eq:div:traceless}) into Eq.~(\ref{eq:perturbed EoM in F(G)}), we obtain
\begin{equation}\label{eq:perturbation equation for F(G)}
\frac{M_{\rm pl}^2}{4}\left[\square h^T_{\mu\nu}-\frac{\tilde{R}}{6}h^T_{\mu\nu} \right]-\frac{\tilde{R}}{3}\left[\nabla_\mu\nabla_\nu-\frac{m_{F(\mathcal{G})}^2}{4}\tilde{g}_{\mu\nu}\right]F''(\tilde{\mathcal{G}})\delta\mathcal{G}=0,
\end{equation}
where $m_{F(\mathcal{G})}^2$ is defined by
\begin{equation}\label{eq:def of mass in F(G)}
m_{F(\mathcal{G})}^2=\frac{{3M_{\rm pl}^2}/{2}}{F''(\tilde{\mathcal{G}})\tilde{R}^2}-\frac{1}{3}\tilde{R}.
\end{equation}
The wave equation for the scalar mode fluctuation is obtained by contractions of Eq.~(\ref{eq:perturbation equation for F(G)}) by $\tilde{g}^{\mu\nu}$,
\begin{equation}\label{eq:mass of F(G)}
[\square-m_{F(\mathcal{G})}^2]\delta \Phi=0.
\end{equation}
where we identify $F''(\tilde{\mathcal{G}})\delta \mathcal{G}$ with the scalar mode fluctuation, $\delta\Phi$,
\begin{align*}
F''(\tilde{\mathcal{G}})\delta \mathcal{G} \equiv \delta\Phi.
\end{align*}
The background curvature dependence of the scalar mode mass is given by Eq.~(\ref{eq:def of mass in F(G)}).

The rest part of Eq.~(\ref{eq:perturbation equation for F(G)}) corresponds to the wave equation for the tensor modes,
\begin{equation}\label{eq:wave Eq for tensor in F(G)}
\square h^T_{\mu\nu}-\frac{\tilde{R}}{6}h^T_{\mu\nu}=0.
\end{equation}
It is more convenient to normalize the traceless parts as $h^T_{ij}=a^2 e_{ij}$.
Then the wave equations for the tensor modes are written as
\begin{equation}\label{eq:wave Eq for tensor in F(G)2}
 \left(-\partial_{0}^2-3H\partial_{0}+\sum_{k}\frac{\partial_{k}^2}{a^2} \right) e_{ij}=0,
\end{equation}
where $H$ is Hubble rate defined as $H\equiv\dot{a}/a$. 
To derive Eq.~(\ref{eq:wave Eq for tensor in F(G)2}) we use $\tilde{R}=12H^2$.

Thus we conclude that a massive scalar mode appears as an additional degree of freedom and the tensor modes propagate with the speed of light under the De Sitter background in $F(\mathcal{G})$ gravity. 
It should be noticed that the perturbation of Gauss-Bonnet invariant (\ref{eq:perturbation of gauss-bonnet}) vanishes and only the tensor modes propagate in a flat spacetime~\cite{Bogdanos:2009tn}.

\section{GWs in $F(R,\mathcal{G})$ gravity\label{GWs in F(R,G)}}
Here we consider $F(R,\mathcal{G})$ gravity, a more general class of theories of the modified gravity.
The action of $F(R,\mathcal{G})$ gravity is defined as an integral of a function of Ricci scalar $R$ and Gauss-Bonnet invariant $\mathcal{G}$~\cite{DeLaurentis:2015fea},
\begin{equation}
S_{RG}=\int d^4x \sqrt{-g}\left(\frac{M_{\rm pl}^2}{2}F(R,\mathcal{G})+\mathcal{L}_{matter}\right).
\end{equation} 
$F(R)$ and $F(\mathcal{G})$ gravity are regarded as a special cases of $F(R,\mathcal{G})$ gravity.
The EoM is obtained by taking the variation of this action with respect to the metric tensor,
\begin{align}\label{eq:EoM F(RG)} \nonumber
&R_{\mu\nu}F_R(R,\mathcal{G})+\frac{1}{2}g_{\mu\nu}\mathcal{G}F_{\mathcal{G}}(R,\mathcal{G})-\frac{1}{2}g_{\mu\nu}F(R,\mathcal{G}) 
\\ \nonumber
&+[g_{\mu\nu}\square-\nabla_\mu\nabla_\nu]F_R(R,\mathcal{G})
\\ \nonumber
&+[2g_{\mu\nu}R\square-2R\nabla_\mu\nabla_\nu-4R_{\mu\nu}\square-4g_{\mu\nu}R^{\rho\sigma}\nabla_\rho\nabla_\sigma
\\ 
&+4{R^\rho}_\mu\nabla_\rho\nabla_\nu+4{R^\rho}_\nu\nabla_\rho\nabla_\mu+{{{R_\mu}^\rho}_\nu}^\sigma\nabla_\rho\nabla_\sigma]F_{\mathcal{G}}(R,\mathcal{G})=0.
\end{align}
where we write
\begin{align}
F_{R}(R,\mathcal{G})\equiv \frac{\partial F(R,\mathcal{G})}{\partial R},
\quad
F_{\mathcal{G}}(R,\mathcal{G})\equiv \frac{\partial F(R,\mathcal{G})}{\partial \mathcal{G}}.
\end{align}

It is known that the degrees of freedom in $F(R,\mathcal{G})$ gravity decrease from four to three under FLRW and De Sitter background~\cite{DeFelice:2010hg}.
Following the procedure developed in the previous section, we find that the gravitational wave is composed by the massless tensor modes and a massive scalar mode.
To achieve the scalar mode mass we evaluate the first order perturbations of $F_R(R,\mathcal{G})$ and $F_\mathcal{G}(R,\mathcal{G})$ around the De Sitter background,
\begin{align}\label{eq:perturbation F(R,G)1}
&F_R(R,\mathcal{G}) \simeq F_{R}(\tilde{R},\tilde{\mathcal{G}})+F_{RR}(\tilde{R},\tilde{\mathcal{G}})\delta R+F_{R\mathcal{G}}(\tilde{R},\tilde{\mathcal{G}})\delta\mathcal{G},
\\ \label{eq:perturbation F(R,G)2}
&F_{\mathcal{G}}(R,\mathcal{G}) \simeq F_{\mathcal{G}}(\tilde{R},\tilde{\mathcal{G}})+F_{\mathcal{G}R}(\tilde{R},\tilde{\mathcal{G}})\delta R+F_{\mathcal{G}\mathcal{G}}(\tilde{R},\tilde{\mathcal{G}})\delta \mathcal{G}.
 \end{align}
Using the relation $\delta \mathcal{G}=\tilde{R}\delta R/3$, Eqs.~(\ref{eq:perturbation F(R,G)1}) and (\ref{eq:perturbation F(R,G)2}) are rewritten as
\begin{align}\label{eq:perturbation F(R,G)3}
&F_R(R,\mathcal{G}) \simeq F_{R}(\tilde{R},\tilde{\mathcal{G}})+\left(1+\frac{\tilde{R}}{3}F_{R\mathcal{G}}(\tilde{R},\tilde{\mathcal{G}}) \right)\frac{\delta \Phi}{F_{RR}(\tilde{R},\tilde{\mathcal{G}})},
\\ \label{eq:perturbation F(R,G)4}
&F_{\mathcal{G}}(R,\mathcal{G}) \simeq F_{\mathcal{G}}(\tilde{R},\tilde{\mathcal{G}})+\left(F_{R\mathcal{G}}(\tilde{R},\tilde{\mathcal{G}})+\frac{\tilde{R}}{3}F_{\mathcal{G}\mathcal{G}}(\tilde{R},\tilde{\mathcal{G}}) \right)\frac{\delta \Phi}{F_{RR}(\tilde{R},\tilde{\mathcal{G}})},
\end{align}
where we identify $F_{RR}\delta R$ with the scalar mode fluctuation $\delta\Phi$.

The wave equation for the scalar model is obtained by the contraction of Eq.~(\ref{eq:EoM F(RG)}) with $\tilde{g}^{\mu\nu}$,
\begin{equation}\label{eq:mass of F(R,G)}
[\square-m_{F(R,\mathcal{G})}^2]\delta \Phi=0,\quad m_{F(R,\mathcal{G})}^2=\frac{F_R-F_{RR}\tilde{R}-\frac{2\tilde{R}^2}{3}F_{R\mathcal{G}}-\frac{\tilde{R}^3}{9}F_{\mathcal{G}\mathcal{G}}}{3F_{RR}+2\tilde{R}F_{R \mathcal{G}}+\frac{\tilde{R}^2}{3}F_{\mathcal{G}\mathcal{G}}}.
\end{equation}
The scalar mode mass depends on the background curvature through Eq.~(\ref{eq:mass of F(R,G)}).
It becomes a generalization from the mass function for $F(R)$ and $F(G)$ gravity.
For $\mathcal{G}=0$ Eq.~(\ref{eq:mass of F(R,G)}) reduces to the wave equation of $F(R)$ gravity (\ref{eq:mass of F(R)}).
Eq.~(\ref{eq:mass of F(R,G)}) coincides with Eq.~(\ref{eq:mass of F(G)}) for a special case $F(R,\mathcal{G})=R+F(\mathcal{G})/M_{\rm{pl}}^2$.

\section{Scalar mode mass\label{Calculation}}
\subsection{Exponential model}
\begin{figure}[ht]
 \centerline{
  \includegraphics[scale=0.6]{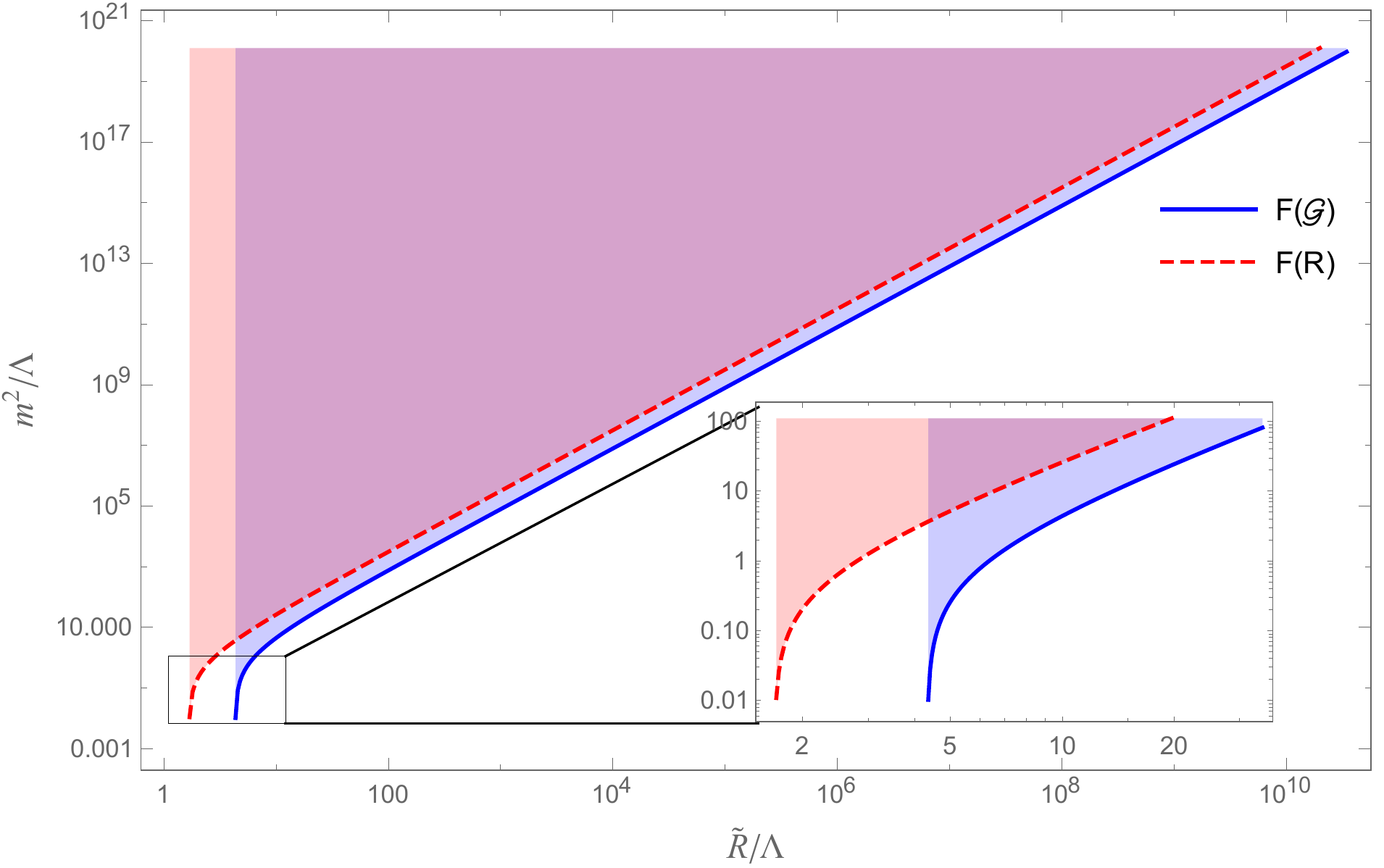}}
  \caption{Behavior of mass-squared of the scalar mode in the exponential models of $F(R)$ and $F(\mathcal{G})$ gravity. 
  The lower limit of mass-squared is plotted by solid and dashed lines as a function of the background curvature.
  Filled areas show the possible regions in each model. 
  \label{Fig:mass in exponential model}}
\end{figure}
De Sitter metric satisfies the Friedmann equation if the energy density of the universe is asymptotically constant.
It gives the simplest background to exhibit the accelerating expansion of the universe.
The exponential model of $F(R)$ gravity is introduced to turn on the cosmological constant for a strong curvature $R>R_0$ and suppress it at the flat limit, $R\rightarrow 0$~\cite{Cognola:2007zu}.
The model is defined by 
\begin{equation}\label{eq:exponential model in F(R)}
F(R)=R-2\Lambda (1-e^{-R/R_{0}}).
\end{equation}
To describe the current accelerating expansion of the universe $R_0$ is fixed smaller than a typical curvature of the universe
and $\Lambda$ is at the present scale of the cosmological constant, $(10^{-33} \rm{eV})^2$. 
From the formula (\ref{eq:mass of F(R)}) the mass for the scalar mode propagation is estimated as 
\begin{equation}\label{eq:mass in F(R) exp}
m_{F(R)}^2=\frac{1}{3}\left(\frac{{R_{0}}^2}{2\Lambda}e^{\tilde{R}/R_{0}} -R_{0}-\tilde{R}\right),
\end{equation}

A similar model is studied in the modified Gauss-Bonnet gravity to switch on the cosmological constant term for $\mathcal{G}>\mathcal{G}_0$~\cite{Odintsov:2018nch},
\begin{equation}\label{model:expF(G)}
F(\mathcal{G}) = -\frac{M_{\rm pl}^2}{2}2\Lambda (1-e^{-{\mathcal{G}}/{{\mathcal{G}_0}}}).
\end{equation}
Substituting Eq.~(\ref{model:expF(G)}) into Eq.~(\ref{eq:def of mass in F(G)}), we compute the mass of the scalar mode propagation,
\begin{equation}\label{eq:mass in F(G) exp}
m_{F(\mathcal{G})}^2=\frac{{\mathcal{G}_0}^2}{4\tilde{\mathcal{G}}\Lambda}e^{\tilde{\mathcal{G}}/\mathcal{G}_0}-\frac{1}{3}\tilde{R}.
\end{equation}
The mass is sensitive to the model parameters $R_0$ and $\mathcal{G}_0$, since the dominant contributions in Eq.~(\ref{eq:mass in F(R) exp}) and Eq.~(\ref{eq:mass in F(G) exp}) come from the exponential terms. 

Differentiating Eqs.~(\ref{eq:mass in F(R) exp}) and (\ref{eq:mass in F(G) exp}) with respect to the constant parameters, $R_0$ and $\mathcal{G}_0$, we find the lower bound of the mass at
\begin{align*}
&\tilde{R}=R_0\left(2+W\left(-\frac{2\Lambda}{e^2R_0}\right)\right),
\\
&\tilde{\mathcal{G}}=2\mathcal{G}_0,
\end{align*}
where $W(x)$ is the Lambert W function.
In Fig.~\ref{Fig:mass in exponential model}, we plot the lower bound of the scalar mode mass-squared as a function of the background curvature. 
The possible regions are shown by filled areas for each model. 
As is shown in Fig.~\ref{Fig:mass in exponential model}, the lower bound monotonically increases for a positive $\tilde{R}$. 
The scalar mode obtains heavy mass and decouples from low energy phenomena at the large curvature, $\Lambda\ll\tilde{R}$, $\Lambda^2\ll\tilde{\mathcal{G}}$.
It should be noticed that the mass-squared becomes negative and the tachyonic mode appears for a small curvature
(See the enlarged plot in Fig.~\ref{Fig:mass in exponential model}). 
Since the modified term vanishes at the limit, $R\rightarrow 0$, the scalar mode disappears.

If we set the model parameters $R_0$ and $\mathcal{G}_0$ near the cosmological constant scale, $\tilde{R}\sim R_0 \sim \Lambda$, 
the scalar mode acquires the mass at the scale, $m^2\sim \Lambda \sim (10^{-33}\, \rm{eV})^2$.
The correlation length of the scalar mode is so long that there is a chance to detect the extra mode beyond GR.
As is shown in Fig.~\ref{Fig:mass in exponential model}, similar behavior is observed in each model and the mass ratio is of order unity, $m^2_{F(\mathcal{G})}/m^2_{F(R)}\sim \mathcal{O}(1)$. 
Thus we conclude that it is hard to test the difference between $F(R)$ and $F(\mathcal{G})$ gravity by the observation of GWs.

\subsection{Power-law model}
Another familiar model of $F(R)$ gravity is the power-law model~\cite{Elizalde:2010ts,Codello:2014sua,Ben-Dayan:2014isa,Rinaldi:2014gua,Motohashi:2014tra,Liu:2018htf}, 
\begin{equation}\label{eq:r^a model}
\mathcal{L}=\frac{M_{\rm pl}^2}{2}\left(R+|R|\times\left| \frac{{R}}{r_0}\right|^{\alpha-1}\right)\quad (\alpha >1),
\end{equation}
where $r_0$ is a constant parameter with mass dimension $\mbox{dim}[r_0]=\rm{eV}^2$.
It is considered that the higher order term of $R$ induces the accelerated expansion at inflation era.
The term decreases more rapidly than Einstein-Hilbert term, then the universe exit from the era.
The modified term gives a negligible contribution for a small curvature and the model is able to pass all the constraints on the solar system.
From Eq.~(\ref{eq:mass of F(R)}), the mass-squared of the scalar mode is given by
\begin{equation}
m_{F(R)}^2=\frac{\tilde{R}}{3\alpha(\alpha-1)}\left[\left(\frac{\tilde{R}}{r_0}\right) ^{1-\alpha}+\alpha(2-\alpha)\right].
\end{equation}
The model is extended to the modified Gauss-Bonnet gravity~\cite{Nojiri:2005am,{Carloni:2017ucm},{Zhong:2018tqn}},
\begin{equation}\label{eq:g^b model}
\mathcal{L}=\frac{M_{\rm pl}^2}{2}\left(R+ \frac{\left|\mathcal{G}\right|^\beta}{{|g_0|}^{\beta-\frac{1}{2}}}\right)\quad (\beta >1),
\end{equation}
where $g_0$ is a constant parameter with the mass dimension, $\mbox{dim}[g_0]=\rm{eV}^4$. 
From Eq.~(\ref{eq:def of mass in F(G)}) we obtain the mass-squared of the scalar mode under the De Sitter background,
\begin{equation}\label{eqs:g^bmodel mass}
m_{F(\mathcal{G})}^2=\frac{{g_0}^{\frac{1}{2}}}{2\beta(\beta-1)}\left| \frac{\tilde{\mathcal{G}}}{g_0}\right|^{1-\beta}-\frac{1}{3}\tilde{R}.
\end{equation}
\begin{figure}[pt]
\centerline{\includegraphics[scale=0.6]{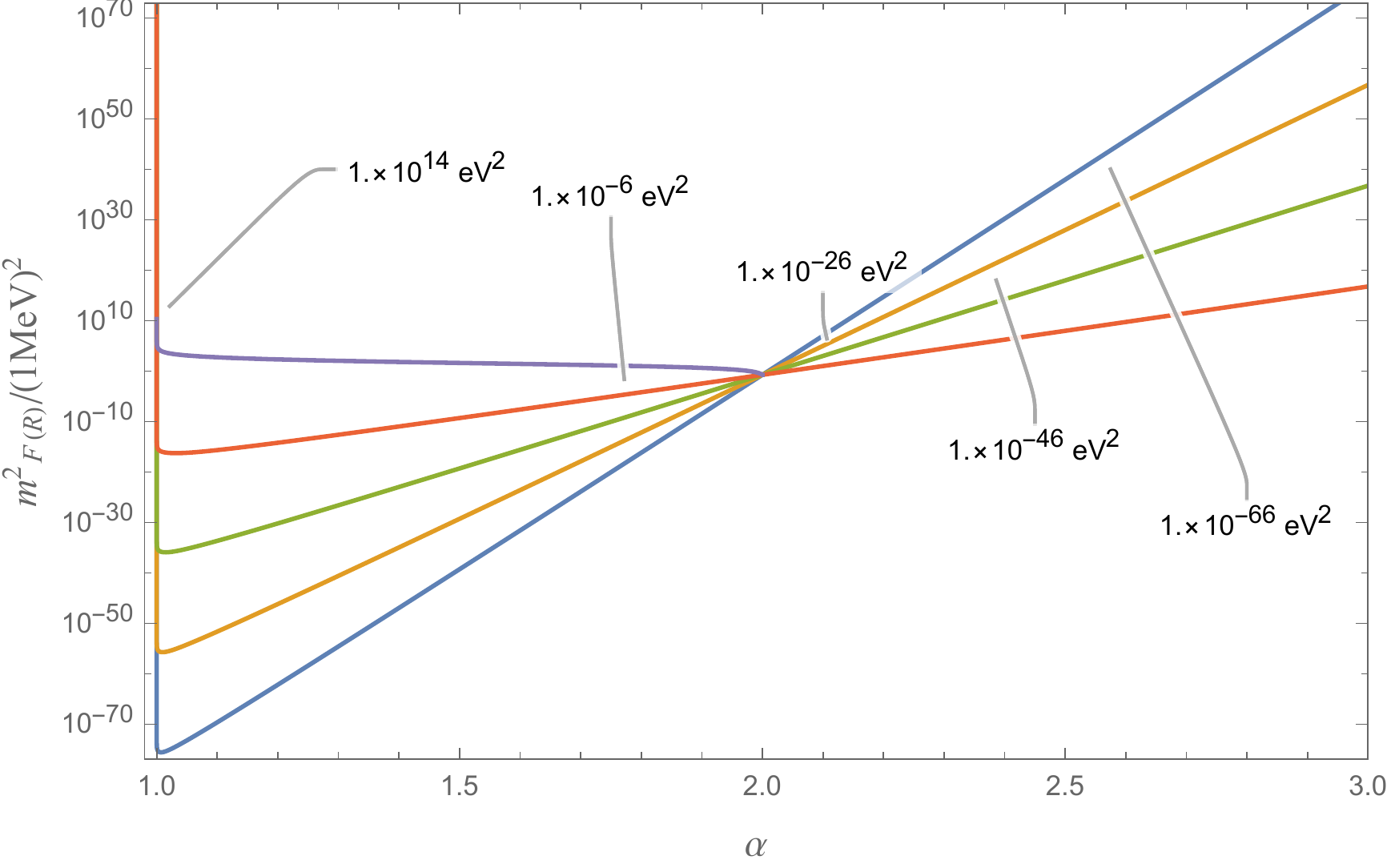}}

\caption{Behavior of mass-squared of the scalar mode in the power-law models of $F(R)$ gravity
for $r_0=(1$$\rm{MeV})^2$, $\tilde{R}=10^{-66},10^{-46},10^{-26},10^{-6},10^{14}$[$\rm{eV}^2$].}
  \label{Fig:mass of r^a model}
\end{figure}
 \begin{figure}[pt]
    \centerline{\includegraphics[scale=0.6]{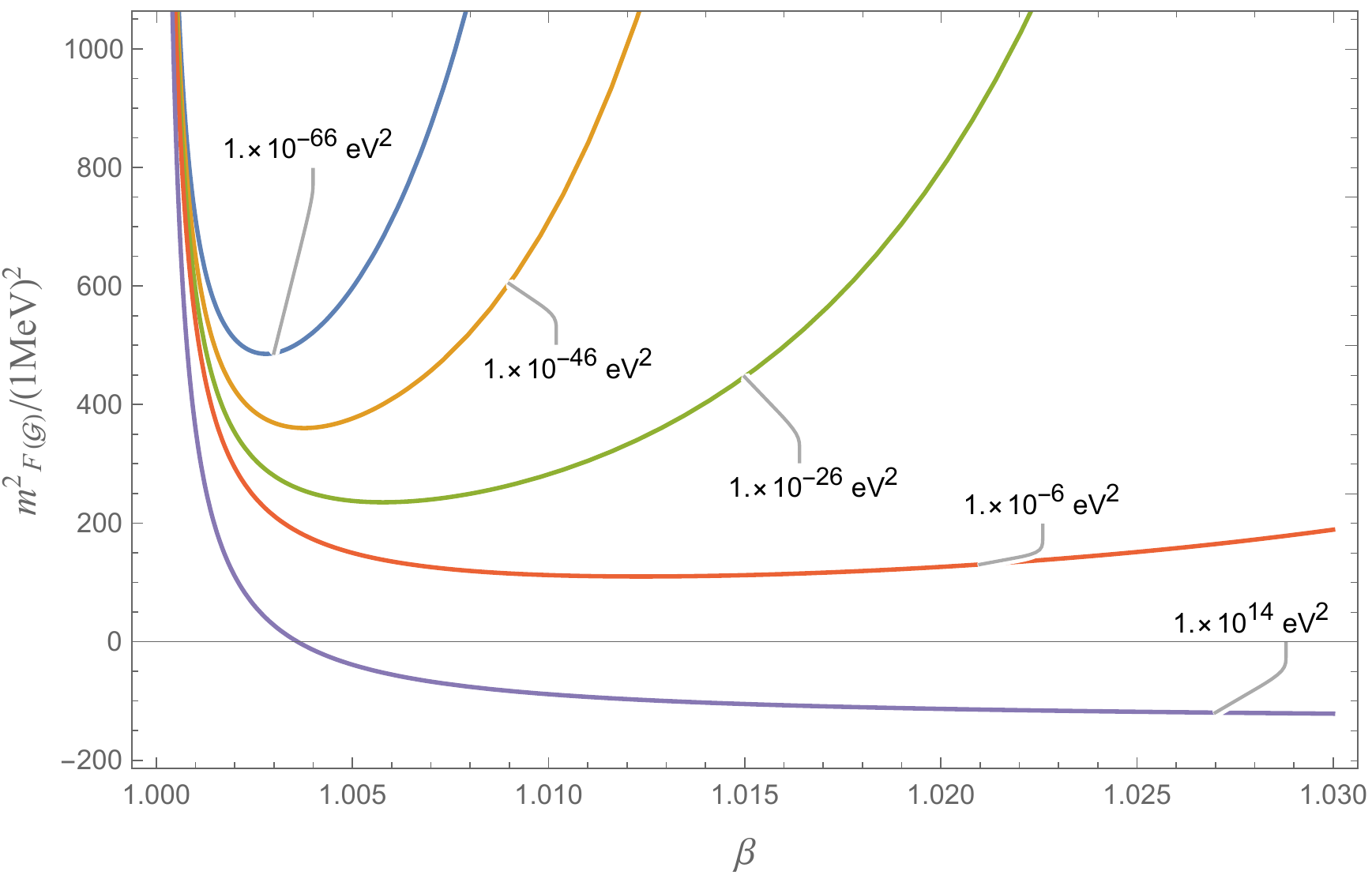}}
  \caption{Behavior of mass-squared of the scalar mode in the power-law models of $F(\mathcal{G})$ gravity
for $g_0=(1\rm{MeV})^4$, $\tilde{R}=10^{-66},10^{-46},10^{-26},10^{-6},10^{14}$[$\rm{eV}^2$].}
\label{Fig:mass of g^b model}

\end{figure}
In Figs.~\ref{Fig:mass of r^a model} and \ref{Fig:mass of g^b model}, we show the behavior of the scalar mode mass as a function of the exponent $\alpha$ or $\beta$. If the background values are smaller than the constant parameters, $\tilde{R}<r_0$ or $\tilde{\mathcal{G}}<g_0$,
we observe a minimum of the scalar mode mass at
\begin{align*}
&\alpha=1-\frac{1}{2\log {\tilde{R}/r_0}},\\
&\beta=1-\frac{1}{\log \tilde{\mathcal{G}}/g_0}.
\end{align*}
\begin{table}[b]
\begin{center}
 {\begin{tabular}{c||cc|c}
    &  \multicolumn{2}{|c|}{$\tilde{R},\tilde{\mathcal{G}}< r_0,g_0$} & $\tilde{R},\tilde{\mathcal{G}}> r_0,g_0$
    \\ \hline
    & minimum & $\alpha,\beta  \to \infty$ &  $\alpha,\beta  \to \infty$
    \\ \hline
$m^2_{F(R)}$ &  $-\frac{2}{3}(e^{1/2}-1)\tilde{R}\log\left(\frac{\tilde{R}}{r_0}\right)$&$\infty$&$-\frac{\tilde{R}}{3}$               
   \\   \hline
$m^2_{F(\mathcal{G})}$&$-\frac{e}{2}{g_0}^{\frac{1}{2}}\log\left(\frac{\tilde{\mathcal{G}}}{g_0}\right)-\frac{\tilde{R}}{3} $     &$\infty$&$-\frac{\tilde{R}}{3}$        
\end{tabular}}
\caption{Mass-squared $m^2_{F(R)}$ and $m^2_{F(\mathcal{G})}$}
 \label{table:F(R) and F(G)}
\end{center}
\end{table}
The explicit expressions of the mass-squared at the minimum are given in Table.~\ref{table:F(R) and F(G)}.
For a large background, $\tilde{R}>r_0$ or $\tilde{\mathcal{G}}>g_0$, the mass-squared decreases asymptotically to a negative value $-\frac{\tilde{R}}{3}$, as the exponent $\alpha$ or $\beta$ increases.
Thus the power-law models have a ghost mode. 
In Fig.~\ref{Fig:mass of r^a model} we observe that the mass-squared has a fixed value, $r_0/6$, which is independent of the background curvature for $\alpha=2$.
In Fig.~\ref{Fig:mass of g^b model} no fixed point appears. Therefore we find some differences on the scalar mode mass in the power-law models of $F(R)$ and $F(\mathcal{G})$ gravity.

\section{Observation of GWs\label{observations}}
It is expected that models of the modified gravity can be tested through the observation of GWs. 
The scalar mode of GWs can be detected in the signals from at least three independent 
GW detectors or LISA-like observatories in the case that the direction of the source is known
~\cite{Abbott:2018utx,Gair:2012nm}.
The first test on the scalar mode was done in LIGO-Virgo for GW170814~\cite{Abbott:2017oio}. 
Expanding GW detector network will allow a stronger test for scalar mode~\cite{Hagihara:2019rny}. 
To estimate the scalar mode mass we follow the procedure to detect the difference in the speed between GWs and electromagnetic waves~\cite{{Will:1997bb},Berti:2011jz,Zakharov:2017svs}. 
In Eq.~(\ref{eq:wave Eq for tensor in F(G)2}), the tensor modes of GWs propagate with the speed of light. 
Thus the scalar mode mass can be evaluated by comparing the propagation speed between the massive scalar and the massless tensor modes.
The observed lower bound on the scalar mode mass depends on sensitivity of GWs detectors~\cite{Hayama:2012au}. 
The lower bound of LIGO-Virgo is estimated to be $m \sim10^{-22}{\rm eV}$ for an expected $\mathcal{O}(10)$ signal-to-noise ratio~\cite{Abbott:2017oio,Lee:2010cg}. 
Close white dwarf binaries are detectable with LISA for the frequency range between 
$3{\rm mHz}$ and $6{\rm mHz}$. 
The sensitivity of LISA to all modes is approximately at the same order for low frequencies, 
$f\lesssim \mathcal{O}(10{\rm mHz})$\cite{Tinto:2010hz,Gair:2012nm}. 
Thus the expected lower bound of LISA is $m\sim 6\times10^{-24}{\rm eV}$\cite{Cooray:2003cv}. 

Below we study the detectable parameter space for the exponential and the power-law models in LIGO-Virgo and LISA. 
We assume that the cosmological constant dominates the energy density of the current universe and set the background curvature, 
$\tilde{R}=4\Lambda$, where $\Lambda$ is the current cosmological constant, $\Lambda\sim(10^{-33}{\rm eV})^2$. 
The background of Gauss-Bonnet invariant is estimated from Eq.~(\ref{eqs:backgroud of GB}). Thus we set $\tilde{\mathcal{G}}=8\tilde{\Lambda}^2/3$.

First, we consider the exponential models discussed in the previous section. 
Substituting the background curvature and Gauss-Bonnet invariant into Eqs.~(\ref{eq:mass in F(R) exp}) and (\ref{eq:mass in F(G) exp}), 
we obtain the scalar mode mass in the exponential model of $F(R)$ gravity,
\begin{equation}\label{eq:mass in F(R) exp for current condition}
m_{F(R)}^2=\frac{\Lambda}{3}\left[\frac{1}{2}\left(\frac{{R_{0}}}{\Lambda}\right)^2e^{4\Lambda/R_{0}} -\left(\frac{{R_{0}}}{\Lambda}\right)-4\right].
\end{equation}
and that of $F(\mathcal{G})$ gravity,
\begin{equation}\label{eq:mass in F(G) exp obs}
m_{F(\mathcal{G})}^2=\Lambda\left[\frac{3}{32}\left(\frac{{\mathcal{G}_0}}{\Lambda^2}\right)^2e^{8\Lambda^2/(3\mathcal{G}_0)}-\frac{4}{3}\right],
\end{equation}
\begin{figure}[t]
    \begin{tabular}{c}

     \begin{minipage}{0.5\hsize}
        \centerline{ \includegraphics[width=8cm]{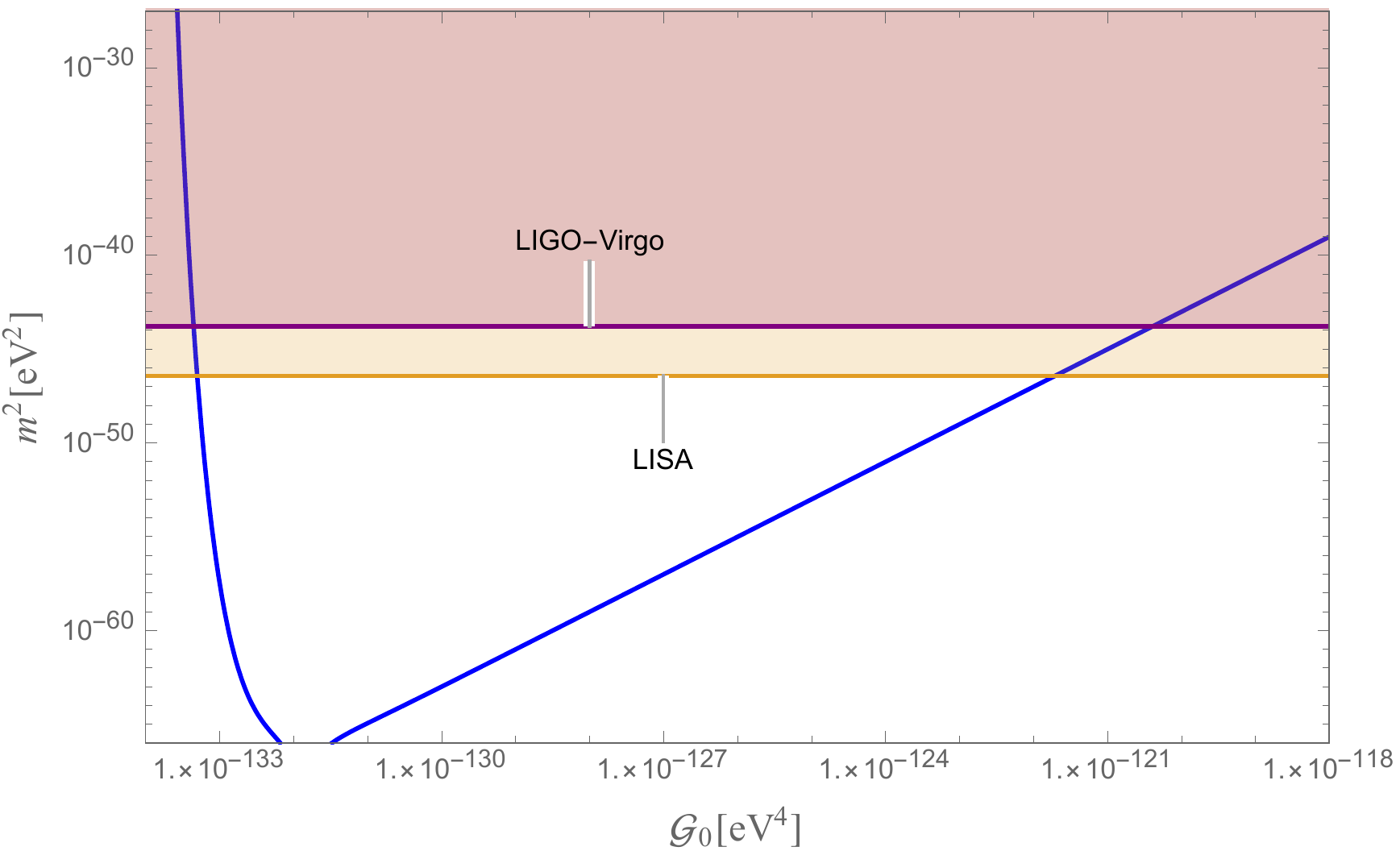} }
          \centerline{(a) $F(\mathcal{G})$ gravity}\hspace{0.6cm}  
     \end{minipage}
     \begin{minipage}{0.5\hsize}
       \centerline{ \includegraphics[width=8cm]{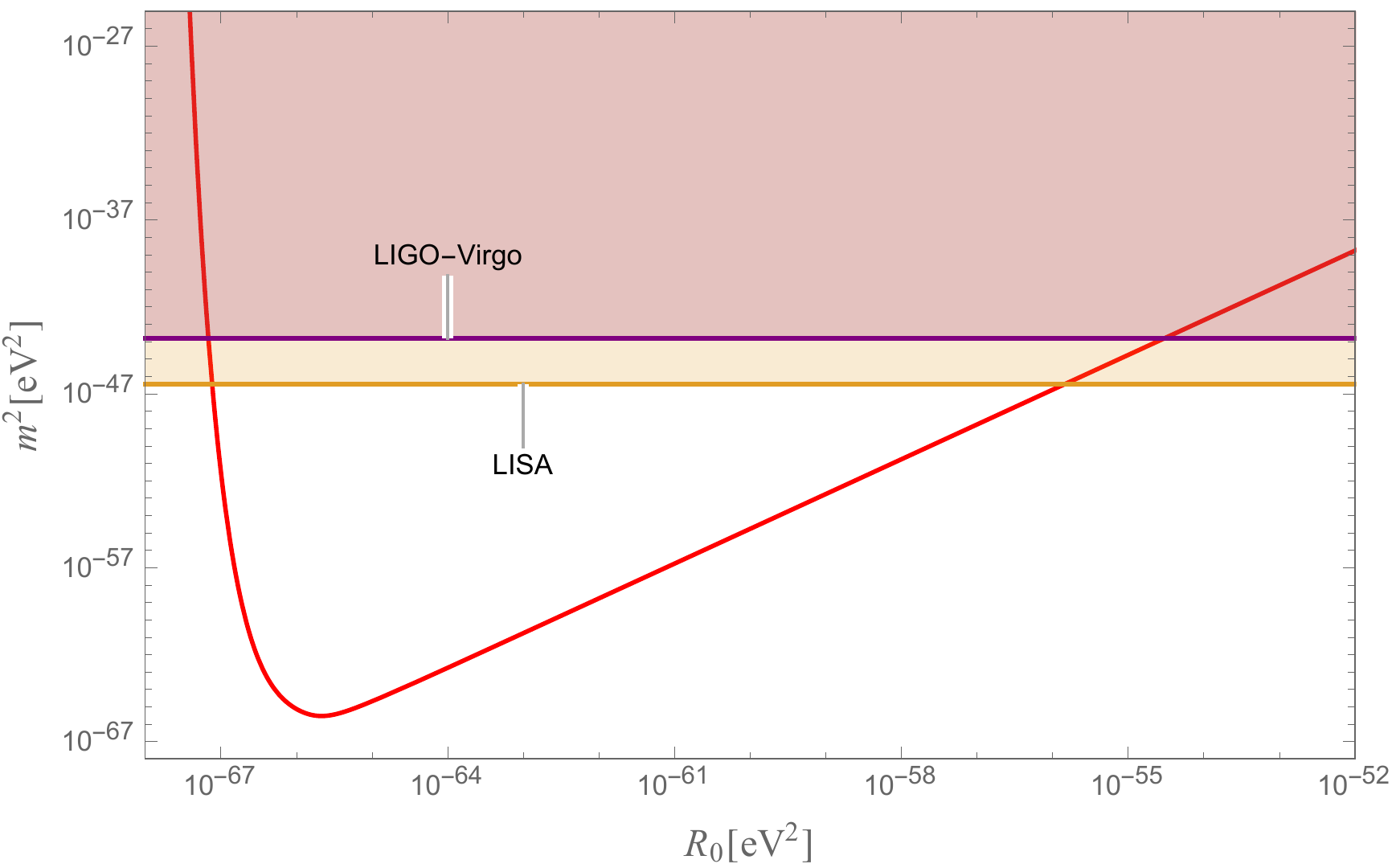}}
            \centerline{(b) $F(R)$ gravity}\hspace{0.6cm}
     \end{minipage}

\end{tabular}
   \caption{Behavior of the scalar mode mass in the exponential models for 
  $\tilde{R}=4\Lambda$ and $\tilde{\mathcal{G}}=8\tilde{\Lambda}^2/3$. 
   The detectable parameter spaces are represented by the filled areas.}
   \label{fig:mass observations in exponential model}
\end{figure}
In Fig.~\ref{fig:mass observations in exponential model}, we plot the mass-squared of the scalar mode mass as a function of parameters, 
$\mathcal{G}_0$ and $R_0$.The detectable parameter spaces of LIGO-Virgo and LISA are shown by the filled areas. 

Because of the exponential blow-up of the scalar mode mass for $R_0 \lesssim \tilde{R}$ in $F(R)$ gravity and 
$\mathcal{G}_0 \lesssim \tilde{\mathcal{G}}$ in $F(\mathcal{G})$ gravity, the difference of the speeds between the scalar and tensor propagation is in the detectable parameter spaces of LIGO-Virgo and LISA. 
As is shown in Fig.~\ref{fig:mass observations in exponential model}, it is not distinguishable in the regions, 
$(10^{-67}{\rm eV}^2)^2 < \mathcal{G}_0 < (10^{-60}{\rm eV}^2)^2$ (LIGO-Virgo), $(10^{-67}{\rm eV}^2)^2 < \mathcal{G}_0 < (10^{-61}{\rm eV}^2)^2$ (LISA) for $F(\mathcal{G})$ gravity 
and $10^{-67}{\rm eV}^2 < R_0 < 10^{-55}{\rm eV}^2$ (LIGO-Virgo), $10^{-67}{\rm eV}^2 < R_0 < 10^{-56}{\rm eV}^2$ (LISA) for $F(R)$ gravity. 
In both exponential models we can tune the parameters, 
$\mathcal{G}_0$ and $R_0$, to reproduce the observed speed of GWs.

Next, we consider the power-law models in $F(\mathcal{G})$ and $F(R)$ gravity. 
Substituting the background curvature $\tilde{R}=4\Lambda$ and Gauss-Bonnet invariant $\tilde{\mathcal{G}}=8\tilde{\Lambda}^2/3$ with the current cosmological constant, 
$\Lambda\sim(10^{-33}{\rm eV})^2$, into the minimum mass in Table.~\ref{table:F(R) and F(G)}, the minimum scalar mode mass is given by a function of the model parameters, ${r_0}$ and $g_0$.

\begin{figure}[t]
    \begin{tabular}{c}

      \begin{minipage}{0.5\hsize}
         \centerline{ \includegraphics[width=8cm]{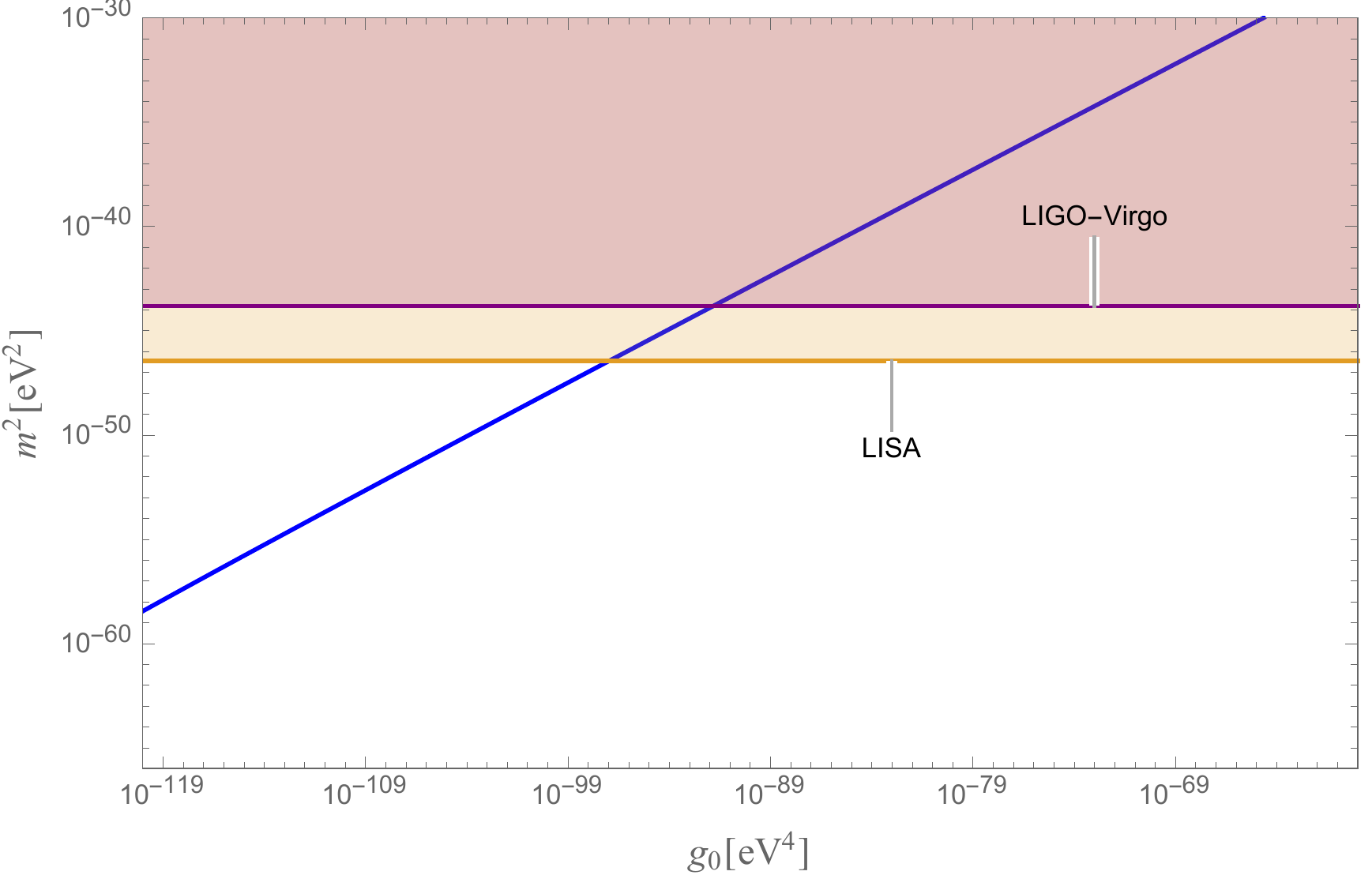}}
            \centerline{(a) $F(\mathcal{G})$ gravity} \hspace{0.6cm}
       \end{minipage}
      \begin{minipage}{0.5\hsize}
         \centerline{ \includegraphics[width=8cm]{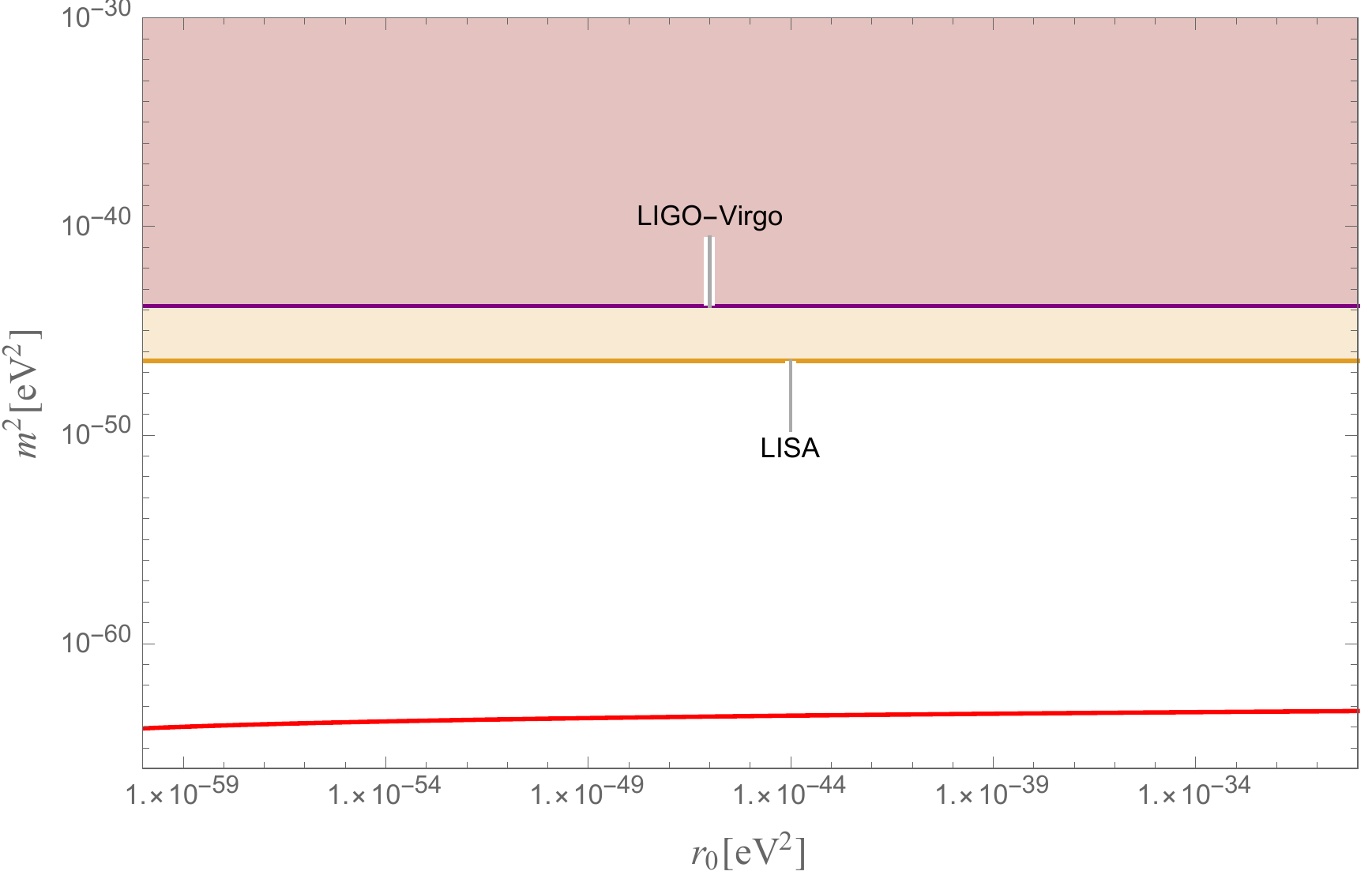}}
            \centerline{(b) $F(R)$ gravity}\hspace{0.6cm} 
      \end{minipage}

 \end{tabular}
    \caption{Behavior of the minimum of the scalar mode mass in the power-law models 
    for $\tilde{R}=4\Lambda$ and $\tilde{\mathcal{G}}=8\tilde{\Lambda}^2/3$. 
    The detectable parameter spaces are represented by the filled areas.}
    \label{fig:mass observations in power-law model}
\end{figure}

We plot the behavior of the minimum scalar mode mass in Fig.~\ref{fig:mass observations in power-law model}. 
In the power-law model of $F(\mathcal{G})$ gravity the minimum of the mass-squared is proportional to
${g_0}^{1/2}\log(\tilde{\mathcal{G}}/g_0)$ with the model parameter $g_0$ in Eq.~(\ref{eq:g^b model}).
The exponent, $1/2$, generates the inclination of the solid line in Fig.~\ref{fig:mass observations in power-law model} (a). 
The scalar mode acquires a larger mass as $g_0$ increases. 
Thus the scalar mode in the power-law model of $F(\mathcal{G})$ gravity is heavy enough to detect if the parameter $g_0$ is greater than 
$\sim (10^{-23}{\rm eV})^4$ in LIGO-Virgo and $\sim (10^{-24}{\rm eV})^4$ in LISA. 
Different behavior is observed for the power-law model of $F(R)$ gravity~(\ref{eq:r^a model}). 
The minimum of the mass-squared is proportional to $\tilde{R}\log(\tilde{R}/r_0)$. 
It is observed that the minimum is nearly independent of the parameter, $r_0$ and much smaller than the detectable parameter spaces by LIGO-Virgo and LISA.

In Table.~\ref{table:obs mass} we take the parameters $g_0$ and $r_0$ at a typical scale of inflation, $10^{16}{\rm GeV}=10^{25}{\rm eV}$ and a low energy scale, 
$10^{-25}{\rm eV}$ and calculate the minima of the mass in each model. 
In the $F(\mathcal{G})$ model the scalar mode develops about 10 times heavier mass than the parameter, $g_0$.
If the parameter, $g_0$, has a typical inflationary scale, the scalar mode decays to other matter during inflation.
Thus it is difficult to test the $F(\mathcal{G})$ model through the scalar mode in the primordial gravitational waves.
Some specific mechanism to stabilize the scalar mode is necessary to regard the scalar mode as an inflaton field which dominates the energy density of the universe at the inflation era.
At a low energy scale, it is expected that the scalar mode mass determine a range of the parameter in the power-law model of $F(\mathcal{G})$ gravity.
\begin{table}[h]
\begin{center}
{\begin{tabular}{c||c|c||c|c|}
 & \multicolumn{2}{|c||}{$F(\mathcal{G})$ gravity} & \multicolumn{2}{|c|}{$F(R)$ gravity}
\\ \hline\hline
$g_0 (F(\mathcal{G})) ,r_0 (F(R))$ & $(10^{25}\rm{eV})^4$ & $(10^{-25}\rm{eV})^4$ &  $(10^{25}\rm{eV})^2$ & $(10^{-25}\rm{eV})^2$
\\ \hline
Minimum mass~[eV]&$2.7\times10^{26}$ &$9.9\times10^{-25}$ & $4.3\times10^{-32}$& $1.6\times10^{-32}$
\end{tabular}}
 \caption{Minima of the squared scalar mode mass}
\label{table:obs mass}
\end{center}
\end{table}

\section{Conclusion\label{conclusion}}
We have investigated the GWs in modified Gauss-Bonnet gravity which is introduced as a generalization of GR.
$F(\mathcal{G})$ and $F(R,\mathcal{G})$ gravity are defined as a family of theories with an arbitrary function of the Gauss-Bonnet invariant in the Lagrangian.

In the theories GWs have extra degrees of freedom and propagate as the massless tensor modes and a massive scalar mode under the De Sitter background.
Evaluating the wave equation in $F(\mathcal{G})$ and $F(R,\mathcal{G})$ gravity, we find formulae to calculate the scalar mode mass.
In Table.~\ref{table:graviton mass} we summarize the features of GWs in the theories under consideration. 
\begin{table}[ph]
\begin{center}
{\begin{tabular}{c||c|c}
Theory & Physical modes & Scalar mode mass
\\ \hline\hline
GR &tensor modes &-
\\ \hline
$F(R)$ &tensor +scalar modes& Eq.~\ref{eq:mass of F(R)}
\\ \hline
$F(\mathcal{G})$&tensor + scalar modes & Eq.~\ref{eq:def of mass in F(G)}
\\ \hline
$F(R,\mathcal{G})$& tensor + scalar modes & Eq.~\ref{eq:mass of F(R,G)}
\end{tabular}}
\caption{GWs in GR, $F(R)$, $F(\mathcal{G})$ and $F(R,\mathcal{G})$ gravity}
\label{table:graviton mass}
\end{center}
\end{table}
We have considered two types of realistic models which can pass all the constraints on the solar system. 
The exponential models are studied at the current cosmological constant scale, $\Lambda \sim(10^{-33}\, \rm{eV})^2$. 
In the models the mass-squared of the scalar mode is of order of the cosmological constant, $O((10^{-33}\, \rm{eV})^2)$.
The mass is small enough to observe the signal of the scalar mode. Similar behavior is observed in the exponential models of $F(R)$ and $F(\mathcal{G})$ gravity. 
On the other hand some different properties are found in the power-law models. The scalar mode mass is fixed independent of the background curvature in $R^2$ gravity. 

To obtain the detectable model parameters by LIGO-Virgo and LISA we have evaluated the model parameter dependence of the scalar mode mass in each model. 
The detectable parameter spaces of LIGO-Virgo and LISA is restricted in the exponential models of $F(R)$ and $F(\mathcal{G})$ gravity. 
It is found that the scalar mode mass is measurable for a wide range of the model parameters in the power-law model of $F(\mathcal{G})$ gravity. 
The study to constrain modified gravity theories has been carried out by using solar system experiments~\cite{DeFelice:2009aj}, the N-body simulations and some observations of the large-scale structure~\cite{Cataneo:2014kaa}.
Only the models of modified gravity that satisfy all the constraints are worth to investigate as a candidate for the theory of gravity.
The measurement of the scalar mode mass with the future GW observations gives complementary information and is expected to improve the current constraints on modified gravity theories.

The present work is restricted mostly to the analysis under the cosmological background, the longest section that GWs propagate. 
We are interested in including the contribution from the spherical symmetric background~\cite{DeFelice:2009aj} and matter effect~\cite{Bamba:2018cup}. 
The chameleon mechanism might affect GWs propagation in the solar system~\cite{Ito:2009nk}. 
It is also interesting to apply the result to the analysis of GWs in $F(T)$ gravity~\cite{Cai:2018rzd} and several modified theories~\cite{Capozziello:2019klx}. 
We will continue our work further and hope to report on these problems.
 
\section*{Acknowledgments}
The authors would like to thank T. Katsuragawa, Y. Matsuo, H. Sakamoto, H. Shimoji and Y. Sugiyama for valuable discussions.

\end{document}